\documentclass[12pt]{iopart}
\usepackage{graphicx}
\usepackage[caption=false]{subfig}
\usepackage{epstopdf}
\usepackage{amsfonts}

\begin{document}

   \title{Hamilton symmetry in relativistic Coulomb systems}

   \author{Uri Ben-Ya'acov}

   \address{School of Engineering, Kinneret Academic College on
   the Sea of Galilee, \\   D.N. Emek Ha'Yarden 15132, Israel}

   \ead{uriby@kinneret.ac.il}

    (\today)

\vskip 2.0cm

\begin{abstract}
Hamilton's hodograph method geometrizes, in a simple and very elegant way, in velocity space, the full dynamics of classical particles in $1/r$ potentials. States of given energy and angular momentum are represented by circular hodographs whose radii depend only on the angular momentum, and hodographs differing only in the energy are related by uniform translations. This feature indicates the existence of an internal symmetry, named here after Hamilton. The hodograph method and the Hamilton symmetry are extended here for relativistic charged particles in a Coulomb field, on the relativistic velocity space which is a 3D hyperboloid $H^3$ embedded in a 3+1 pseudo-Euclidean space.

The key for the simplicity and elegance of the velocity-space method is the linearity of the velocity equation, a unique feature of $1/r$ interactions for Newtonian and relativistic systems alike. Although with hodographs much more complicated than for Newtonian systems, the main features of the Hamilton symmetry persist in the full relativistic picture : (1) general hodographs may be represented as linearly displaced base energy-independent circles, (2) hodographs corresponding to same angular momentum but with different energies are connected via translations along geodesics of the velocity space. As an internal symmetry over and beyond central symmetry, the Hamilton symmetry is equivalent to the Laplace-Runge-Lenz symmetry and complements it.
\end{abstract}

\noindent{\it Keywords\/} : {hodograph, relativistic Coulomb system, relativistic velocity space, Hamilton's vector, Hamilton symmetry, rapidity, Laplace-Runge-Lenz symmetry}

\vskip30pt

\eject

\section{Introduction} \label{sec: Intro}

Hodographs are the orbits in velocity space that correspond to the trajectory of a particle in ordinary space, traversed by the tip of the velocity vectors when these are drawn all starting from the same point, the origin of velocity space. The hodograph method -- studying the dynamics of a system in velocity space -- was originally invented by Hamilton  \cite{Hamilton1847} and successfully applied \cite{Maxwell,Goodstein96} to prove geometrically the relation between Kepler's laws and Newton's law of universal attraction. Although largely unfamiliar with, it is a method much simpler and more elegant to study Newtonian Kepler/Coulomb (KC) systems than the familiar analytic solutions in ordinary space \cite{Milnor1983,Sivardiere1992,GonVilla.etal,Butikov2000,Derbes2001,KowenMathur2003,Munoz2003,Carinena.etal2016}.

\begin{flushleft}
\begin{figure}[h]
  \begin{minipage}[b]{295pt}
  \begin{flushleft}
  \caption{\label{fig:11} \textbf{Newtonian Hamilton symmetry} :  $C_o(\ell)$ (doted, red) is the minimum-energy circle, centred at the origin O of the velocity space. The hodograph circle $C(\ell,E')$ (bold line, purple, with the same radius) is centred at M. The Hamilton vector $\vec B_o = \vec{\rm{OM}}$ uniformly translates $C_o(\ell)$ into $C(\ell,E')$, away from the origin. The drawing, with $R < \left|\vec B_o\right|$, corresponds to a bound state with the velocity vector making a whole round. For unbound states with $R > \left|\vec B_o\right|$ the hodograph reduces to a circular arc.}
  \end{flushleft}
  \end{minipage}
  \hskip15pt
  \includegraphics[width=6.5cm]{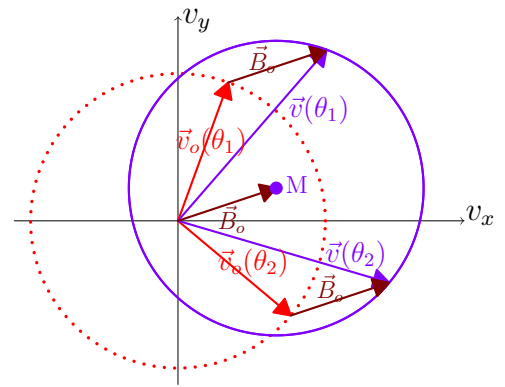}
\end{figure}
\end{flushleft}

With potential $\kappa/r$, the common feature of all Newtonian KC hodographs is them being circles, or circular arcs, with radii depending very simply on the angular momentum $\ell$, $R = |\kappa| / \ell$, independent of the energy; the energy dependence enters only through the magnitude of the (constant) vector which determines the centre of the hodograph circle : As \Fref{fig:11} shows, the hodograph (for given values of total energy $E'$ and angular momentum $\ell$) is the circle $C(\ell,E')$ (bold line) which is just the circle $C_o(\ell)$ (dotted line, centred at the origin of velocity space, corresponding to the minimum energy state) uniformly translated by a (constant) vector $\vec B_o(\ell,E')$.

Since all these hodographs are similar circles, the key for distinguishing between different (energy dependent) configurations is the constant translating vector $\vec B_o$, known as the {\it Hamilton vector}. The existence of a constant or conserved quantity that determines the configuration of a physical system is usually regarded as an indication of a symmetry. Different central vectors correspond to different energies. It is therefore appropriate that these features be regarded as constituting a symmetry, which we call {\it Hamilton symmetry} -- a symmetry that acts on the hodographs in velocity space, its action being the transition between hodographs corresponding to different energy states for the same value $\ell$ of the angular momentum.

The key for the simplicity and elegance of the hodograph method for Newtonian KC systems is the linearity of the velocity equation, which is a unique feature of $1/r$ interactions (Eq. \eref{eq: veqangKC}). The success of the hodograph method for these systems prompts attempting using it also for relativistic charged particles in Coulomb fields. Although the relativistic spatial trajectories are much more complicated \cite{Boyer2004,MunozPavic2006}, it is indeed found that the same key feature -- linearity of the velocity equation -- persists into the relativistic regime, allowing relatively simple analytic discussion (Eqs. \eref{eq: vecueqang} \& \eref{eq: uoeqang}). A thorough account of the relativistic hodograph equations and solutions is contained in a recent publication (\cite{Relhod}; see also \cite{IARD2016} for an exposition of the main results).

Since the Newtonian velocity space is the 3-D Euclidean space ${\cal V}_{\rm N} = \left\{ \vec v \in \mathbb{R}^3\right\}$, the (uniform) hodograph translations are along geodesics of ${\cal V}_{\rm N}$ -- in fact, they are just the Galilei transformations for velocities. This is simple and straight-forward, and doesn't offer much interest. However, on turning to the relativistic picture it becomes intricate and intriguing : The relativistic velocity space (RVS) is an hyperboloid, the space of all future-directed time-like unit 4-vectors ${\cal V}_{\rm rel} \equiv \left\{ u^\mu = \left(\sqrt{1 + \vec u\, ^2} , \vec u \right) \, , \, \vec u \in \mathbb{R}^3\right\}$ embedded in 4-D pseudo-Euclidean space (see \Sref{sec: RVS}). Even with an essentially different geometry, Hamilton symmetry may be established on ${\cal V}_{\rm rel}$ with the same main features as the Newtonian Hamilton symmetry, namely, that
\begin{enumerate}
 \item {General hodographs may be represented as linearly displaced base circular energy-independent hodographs.}
 \item {Translations along geodesics of the velocity space transform between hodographs corresponding to same angular momentum $\ell$ but with different energies.}
\end{enumerate}

The purpose of the present article is to demonstrate how the Hamilton symmetry manifests in relativistic systems. In the first part of the paper, Sections \ref{sec: hamKC}, \ref{sec: RVS} \& \ref{sec: hodrelC}, the Newtonian Hamilton symmetry, the relativistic velocity space and the relativistic hodograph equations are briefly reviewed. Then, in the second, main part, the Hamilton symmetry is established on the RVS : The general energy dependence of relativistic Coulomb hodographs is derived, demonstrating their decomposition into a base hodograph circle and its energy-dependent translation; and it is demonstrated that energy variations of the hodographs are along geodesics of the RVS (Sections \ref{sec: endep} \& \ref{sec: envar}). The nature of the solutions for relativistic Coulomb systems depends on the ratio $|\kappa|/\ell$ which, in the Newtonian limit, is the magnitude of the velocity on the minimum-energy circle $C_o(\ell)$. Therefore, a Newtonian limit exists only for $\ell > |\kappa|$, while for $\ell \le |\kappa|$ the hodographs are exclusively relativistic with distinctive geometrical appearance. The application of the foregoing general results to the geometrical interpretations of the hodographs and their construction relative to ${\cal V}_{\rm rel}$ are therefore discussed distinguishing between the various possible cases (\Sref{sec: geoint}). Finally, the identification of the relativistic Hamilton vector is discussed, together with its relation with the relativistic Laplace-Runge-Lenz (LRL) vector (\Sref{sec: RelHamvec}). Some features of the RVS that are used in the following are listed in the Appendix.

{\it Notation}. The convention $c=1$ is used throughout, unless specified otherwise. Events in Minkowski space-time are $x^\mu = \left(x^0,x^1,x^2,x^3\right)$, with metric tensor $g_{\mu\nu} = {\rm diag} \left(-1,1,1,1\right) \, , \, \mu,\nu = 0,1,2,3$. For any 4-vectors $a^\mu = (a^0,\vec a)$ and $b^\mu = (b^0,\vec b)$, their inner product is then $a \cdot b = -a^0 b^0 + \vec a \cdot \vec b$.

\vskip20pt

\section{The Hamilton symmetry in classical Kepler/Coulomb systems} \label{sec: hamKC}

For a Newtonian particle in a $\kappa/r$ potential, application of angular momentum conservation
 \begin{equation} \label{eq: angmomKC}
 \ell = m r^2 \frac{d\theta}{dt}
 \end{equation}
brings the equation of motion to the form\footnote{With the $x{\rm -}y$ plane as the spatial plane of motion, $\left(r,\theta\right)$ are polar coordinates with unit vectors \[\hat r = \cos\theta \hat x + \sin\theta\hat y \, , \, \hat\theta = -\sin\theta \hat x + \cos\theta \hat y\]}
\begin{equation}\label{eq: veqangKC}
 \frac{d \vec v}{d \theta} = \frac{\kappa}{\ell} \hat r \, ,
\end{equation}
with the immediate integral
 \begin{equation} \label{eq: hodoN}
\vec v = \vec B_o - \frac{\kappa}{\ell} \hat \theta \, ,
 \end{equation}
using the relation $\hat r = - d\hat\theta / d\theta$ and with $\vec B_o$ some arbitrary constant vector. Clearly, the simplicity of equation \eref{eq: veqangKC} which leads to this solution is due to the $1/r^2$ force.

The solution \eref{eq: hodoN} describes a circle (or at least a circular arc for unbound systems) in velocity space,
 \begin{equation} \label{eq: hodoN2}
 \left( \vec v - \vec B_o \right)^2 = \frac{\kappa^2}{\ell^2} \, ,
 \end{equation}
centred around $\vec B_o$ and with radius $|\kappa| / \ell$. Using the relation $v_\theta = r \dot\theta = \ell/(mr)$, the energy integral becomes, for total energy $E\,'$,
\begin{equation}\label{eq: energyintN}
 \frac{m {\vec v \,}^2}{2} + \frac{\kappa}{r} = \frac{m {\vec v \,}^2}{2} + \frac{m \kappa}{\ell} v_\theta = E\,'
\end{equation}
from which it is easily verified, substituting \eref{eq: hodoN} in \eref{eq: energyintN}, that
\begin{equation}\label{eq: BoNewt}
 B_o = \left|\vec B_o\right| = \sqrt{\frac{2E\,'}{m} + \frac{\kappa^2}{\ell^2}} \, .
\end{equation}
The circular hodographs correspond to the conic sections trajectories in ordinary space. The direction of $\vec B_o$ determines the orientation of the spatial trajectory.

The ensuing picture is therefore : The canonical circle $C_o(\ell) = \{\vec v_o(\theta)\}$ with $\vec v_o(\theta) = -(\kappa/\ell) \hat \theta$ is the base hodograph, corresponding to the minimum energy state for a given $\ell$ ($B_o = 0$). A general hodograph is a similar circle $C(\ell,E') = \{\vec v(\theta)\}$ \eref{eq: hodoN} which is the base circle $C_o(\ell)$ uniformly displaced by the constant vector $\vec B_o(E',\ell)$. Since $C_o(\ell)$ is independent of the energy, the energy dependence of the hodograph enters only via the magnitude \eref{eq: BoNewt} of $\vec B_o$.

The vector $\vec B_o$ is known as the {\it Hamilton vector}. Clearly, the transition between any two hodographs corresponding to the same value of the angular momentum is by uniform translations in velocity space, the two hodographs being related by some constant vector which is the difference of their defining Hamilton vectors. Since the Newtonian velocity space is the 3-D Euclidean ($E^3$) space ${\cal V}_{\rm N} = \left\{ \vec v \in \mathbb{R}^3\right\}$, the hodograph translations are the geodesic symmetries of ${\cal V}_{\rm N}$, changing the energy while keeping the same value of angular momentum.

As a final remark, it is noted that an essential aspect of the hodograph method is the change from the usual $t$-dependence of the equations of motion to $\theta$-dependence (Eq. \eref{eq: veqangKC}), employing angular-momentum conservation. The Hamilton symmetry is therefore an extension of central symmetry. Further information and discussions regarding the classical KC analytic hodograph solution may be found in various publications \cite{Milnor1983,Sivardiere1992,GonVilla.etal,Butikov2000,Derbes2001,KowenMathur2003,Munoz2003,Carinena.etal2016}.

\vskip20pt

\section{The relativistic velocity space}\label{sec: RVS}

The relativistic velocity space (RVS) is the space of all future-directed time-like unit 4-vectors, and its usage has the virtue of transforming and displaying kinematical and dynamical properties in a purely geometrical manner. Since there is no explicit dependence on time, different points in it may correspond either to the velocity states of different particles in some instantaneous reference frame, or the velocity state of a particle in different instances along its world-line.

To the author's best knowledge, the most thorough account of the RVS is by Rhodes and Semon \cite{RhodesSemon2004}. They point at an exercise contained already in the 1951 English edition of Landau \& Lifshitz' ``Classical theory of fields" \cite{LLF5} as the first time the idea of RVS appeared in English (for an account of the earlier history of the RVS see a comment by Criado and Alamo \cite{CriadoAlamo2001}). While the Landau \& Lifshitz' exercise dealt only with the metric properties of the RVS, later applications of the RVS \cite{RhodesSemon2004,Urbantke1990,Aravind1997} focus on geometrical derivation of the Thomas-Wigner rotation.

So far, the association of geometry to the RVS used the spatial velocity $\vec v$ for the RVS coordinates, and the metric properties derived from the Lorentz formulae for relativistic velocity addition \cite{LLF5,CriadoAlamo2001}. Instead, we use here the relativistic velocity 4-vector $u^\mu = \left(\gamma\left(v\right), \gamma\left(v\right) \vec v\right)$ with the RVS defined as
\begin{equation}\label{eq: Vrel}
 {\cal V}_{\rm rel} \equiv \left\{ u^\mu = \left(u^0 , \vec u \right) | u^0 = \sqrt{1 + \vec u\, ^2} \right\}
\end{equation}
This is a 3-D unit hyperboloid ($H^3$) embedded in a 4-D pseudo-Euclidian space
\begin{equation}\label{eq: E13}
 E^{(1,3)} = \left\{ w^\mu = \left( w^0,\vec w \right) \in \mathbb{R}^4 | g_{\mu\nu} = {\rm diag} \left(-1,1,1,1\right) \right\}
\end{equation}
Also, for any $u^\mu \in {\cal V}_{\rm rel}$, let ${\cal T}_u\left({\cal V}_{\rm rel}\right) \subset E^{(1,3)}$ be the hyperplane tangent to ${\cal V}_{\rm rel}$ at $u^\mu$. Any vector $A^\mu \in {\cal T}_u\left({\cal V}_{\rm rel}\right)$ is space-like, satisfying $A^2 > 0 , \, A \cdot u = 0$.

Using $u^\mu$ as the coordinates of the RVS allows to relate the geometrical properties of the hyperbolic space with relativistic kinematics and dynamics in a Lorentz-covariant manner. The geometrical properties of the RVS then follow straight-forward. Further properties of the RVS are discussed in the Appendix.

\vskip20pt

\section{The hodograph equations for relativistic Coulomb systems} \label{sec: hodrelC}

A relativistic Coulomb system consists of a point particle with mass $m$ whose dynamics is determined by the Hamiltonian \cite{LLF39}
 \begin{equation} \label{eq: Ham}
 H\left(\vec r, \vec p \right) = \sqrt {{\vec p \,}^2 + m^2}  + \frac{\kappa}{r}
 \end{equation}
with the equation of motion
 \begin{equation} \label{eq: eqmot}
 \frac{d \vec p}{d t} = \frac{\kappa}{r^3} \vec r \, .
 \end{equation}
Solving the momentum equation \eref{eq: eqmot} in the velocity space provides an elegant and simple solution for the motion in a relativistic Coulomb system. The key to the hodograph or velocity space picture is the conservation law for the angular momentum  $\vec \ell  = \vec r \times \vec p$,
 \begin{equation} \label{eq: angmom}
 \ell = \left( E - \frac{\kappa}{r} \right) r^2 \frac{d \theta}{d t} \, ,
 \end{equation}
the relativistic counterpart of Kepler's 2nd law. It allows, as in the original Hamilton's hodograph method, transition to $\theta$ as the hodograph parameter. The (spatial) linear momentum of the particle is $\vec p = m\vec u$, and the conserved energy $E = H\left(\vec r, \vec p \right)$ is
\begin{equation}\label{eq: energyint1}
 m u^0 + \frac{\kappa}{r} = E \, ,
\end{equation}
so the momentum equation \eref{eq: eqmot} becomes
\begin{equation}\label{eq: vecueqang}
 \frac{d \vec u}{d \theta} = \frac{\kappa}{\ell} u^0 \hat r \, .
\end{equation}
Using the polar representation $\vec u = {u_r}\hat r + {u_\theta } \hat\theta$, the polar-angular equations derived from \eref{eq: vecueqang} are
\begin{equation}\label{eq: polangeq}
 \frac{d u_\theta}{d \theta} = -u_r \quad , \quad \frac{d u_r}{d \theta} = \frac{\kappa}{\ell} u^0 + u_\theta \, ,
\end{equation}
complemented by the $u^0$-equation also derived from \eref{eq: vecueqang},
\begin{equation}\label{eq: uoeqang}
 \frac{d u^0}{d \theta} = \frac{\vec u}{u^0} \cdot \frac{d \vec u}{d \theta} = \frac{\kappa}{\ell} \vec u \cdot \hat r = \frac{\kappa}{\ell} u_r \, .
\end{equation}
Equations \eref{eq: vecueqang} (or, alternatively, \eref{eq: polangeq}) and \eref{eq: uoeqang} constitute the relativistic hodograph equations.

The relation
\begin{equation}\label{eq: uthetar}
 u_\theta = \frac{\ell}{m r} \, ,
\end{equation}
another consequence of \eref{eq: angmom}, allows transition from spatial dependencies to velocity dependencies. In particular, using \eref{eq: uthetar} the energy integral \eref{eq: energyint1} becomes
\begin{equation}\label{eq: energyint2}
 u^0 + \frac{\kappa}{\ell} u_\theta = \frac{E}{m}
\end{equation}
which may also be recognized as an immediate integral of \eref{eq: polangeq} and \eref{eq: uoeqang}.

The virtue of equations \eref{eq: polangeq}, \eref{eq: uoeqang} and \eref{eq: energyint2} is their linearity with constant coefficients in the polar representation, a unique feature of the $1/r$ interaction. Therefore, explicit solutions providing the hodographs and the corresponding spatial trajectories are quite immediate to get. These solutions are obtained and discussed in \cite{Relhod,IARD2016}. Here we proceed to present and discuss the relativistic Hamilton symmetry.

\vskip20pt

\section{General energy dependence of the relativistic hodographs} \label{sec: endep}

Recalling that the velocity space ${\cal V}_{\rm rel}$ is embedded in a psudo-Euclidean space $E^{(1,3)}$, the hodograph equations \eref{eq: vecueqang} and \eref{eq: uoeqang} may be combined and considered for arbitrary orbits $w^\mu \left( \theta \right) = \left(w^0, \vec w \right)$ in $E^{(1,3)}$ (not necessarily confined to $\mathcal{V}_{\rm rel}$)  as
\begin{equation}\label{eq: weqgen}
 \frac{d w^\mu}{d \theta} = {\Omega^\mu}_\nu w^\nu
\end{equation}
with
\begin{equation}\label{eq: Omega}
{\Omega^\mu}_\nu = \frac{\kappa}{\ell} \left(\begin{array}{*{20}{c}}
  0 & \vline & {\hat r} \\
\hline
  {\hat r} & \vline & 0
\end{array}
\right) = \frac{\kappa}{\ell} \left(\begin{array}{*{20}{c}}
  0 & \vline & {\cos \theta} & {\sin \theta} & 0 \\
\hline
  {\cos \theta} & \vline & 0 & 0 & 0 \\
  {\sin \theta} & \vline & 0 & 0 & 0 \\
  0 & \vline & 0 & 0 & 0
\end{array}
\right)
\end{equation}
Since $\Omega_{\mu\nu} = -\Omega_{\nu\mu}$, \eref{eq: weqgen} generates rotation in $E^{(1,3)}$, implying constant inner products, so that any solution of the hodograph equations is of constant magnitude, and their motion can only be rotational.

The 4-vector
\begin{equation}\label{eq: vodef}
 v_o^\mu = \left( 1, \vec v_o \right) = \left( 1, - \frac{\kappa}{\ell} \hat \theta \right)
\end{equation}
is a particular solution of \eref{eq: weqgen} which is constant relative to the polar coordinates. Since $\Omega_{\mu\nu}$ has constant coefficients relative to the polar coordinates, $v_o^\mu$ is recognized as the axis vector of this rotation : While it rotates in $E^{(1,3)}$ with the polar coordinate system, every other solution of \eref{eq: weqgen} rotates around it. Its spatial part is the classical 3-vector $\vec v_o = - \left( \kappa/\ell \right) \hat \theta$ which generates the base canonical circle $C_o(\ell)$. $v_o^\mu$ is therefore the $E^{(1,3)}$ extension of $\vec v_o$.

The physical significance of $v_o^\mu$ is that the energy integral \eref{eq: energyint2} is identified, for any hodograph $u^\mu$ in $\mathcal{V}_{\rm rel}$, as the constant $u$-component along $v_o^\mu$,
\begin{equation}\label{eq: energyint}
 u^0 + \frac{\kappa}{\ell} u_\theta = -u \cdot v_o = \frac{E}{m}
\end{equation}
Therefore, while rotation around $v_o^\mu$ keeps the value of the energy $E$, translation along $v_o^\mu$ transforms between hodographs corresponding to different values of the energy. These symmetry relations allow deduction of the generic energy dependence of the hodographs, even without solving the hodograph equations :

A general hodograph is denoted in the following $u^\mu \left( \theta | E, \ell\right)$. $E$ and $\ell$ are fixed parameters, while $\theta$ is the orbit's variable. An infinitesimal variation of the energy $\delta E$ induces the hodograph variation
\begin{equation}\label{eq: hodovar}
 \delta_E u^\mu = u^\mu \left( \theta | E + \delta E, \ell\right) - u^\mu \left( \theta | E, \ell\right) = \frac{\partial u^\mu}{\partial E} \left( \theta |E,\ell \right)\delta E
\end{equation}
On $\mathcal{V}_{\rm rel}$ the constraints $u \cdot \delta_E u = 0$, and by \eref{eq: energyint} also $\delta E = -m \delta_E u \cdot v_o$, imply that
\begin{equation}\label{eq: deludelE}
 \frac{\partial u^\mu}{\partial E} = \frac{E u^\mu - m v_o^\mu}{\Lambda^2 \left( E,\ell \right)}
\end{equation}
with
\begin{equation}\label{eq: defLam}
 \Lambda^2 \left( E,\ell \right) = \left(E u - m v_o\right)^2 = E^2 + m^2 \left(\frac{\kappa^2}{\ell^2} - 1\right) \, .
\end{equation}
The vector $\partial u^\mu / \partial E \in {\cal T}_u\left({\cal V}_{\rm rel}\right)$, therefore a space-like 4-vector. Satisfying the hodograph equations hence with constant magnitude, necessarily $\Lambda^2 \left( E,\ell \right) = \left(\partial u / \partial E\right)^{-2} > 0$.

Regarding \eref{eq: deludelE} as a 1st order ODE in $E$, and using the fact that $v_o^\mu$ is independent of $E$, it is easily integrated. The solution is a linear combination of $v_o^\mu (\theta |\ell)$ with another energy-independent vector $n_o^\mu (\theta |\ell)$ which is also a solution of the hodograph equation \eref{eq: weqgen} and linearly independent of $v_o^\mu$ (its explicit form will be found in the following, depending on $\kappa/\ell$),
\begin{equation}\label{eq: genhod}
 u^\mu \left( \theta |E,\ell \right) = Q(E,\ell) v_o^\mu(\theta |\ell) +  \Lambda\left(E,\ell\right) n_o^\mu (\theta |\ell)
\end{equation}
where
\begin{equation}\label{eq: wforgenhod}
 Q(E,\ell) = \cases{ \frac{E}{m \left( 1 - {\raise0.7ex\hbox{${\kappa^2}$} \!\mathord{\left/
 {\vphantom {{\kappa^2} {\ell^2}}} \right.\kern-\nulldelimiterspace} \!\lower0.7ex\hbox{${\ell^2}$}} \right)} & for $\ell \ne |\kappa|$ \\
 \frac{m}{2E} & for $\ell = |\kappa|$ \\}
\end{equation}
The velocity vector $u^\mu (\theta |E,\ell)$ is therefore a superposition of $v_o^\mu (\theta |\ell)$ and $n_o^\mu (\theta |\ell)$ with coefficients depending very simply on the energy. The result is a combined rotation: Both $v_o^\mu(\theta |\ell)$ and $n_o^\mu(\theta |\ell)$ are solutions of the hodograph equation \eref{eq: weqgen}. They have therefore constant magnitudes and maintain a constant relative angle (in $E^{(1,3)}$), so they may only revolve together. The vector $q^\mu(\theta |E,\ell) \equiv Q(E,\ell) v_o^\mu(\theta |\ell)$ (which is not an hodograph, not lying on $\mathcal{V}_{\rm rel}$) rotates in $E^{(1,3)}$ around the $w^0$-axis, with the vector $B^\mu (\theta |E,\ell) \equiv \Lambda(E,\ell) n_o^\mu (\theta |\ell)$ rotating relative to it and complementing it so that $u^\mu = q^\mu + B^\mu$ is on $\mathcal{V}_{\rm rel}$.

\vskip20pt

\section{Energy variations of the relativistic hodographs are geodesic} \label{sec: envar}

The Newtonian velocity space is the 3-D Euclidean space $\mathcal{V}_{\rm N} = \left\{ \vec v \right\}$. In Newtonian Hamilton symmetry, the hodograph circles are uniformly transformed into other hodographs along straight lines, which are the geodesics of $\mathcal{V}_{\rm N}$. Very interestingly, although the relativistic hodographs are much more complicated than the Newtonian ones, these features persist into the relativistic realm. We now show that the energy variations of all the hodographs $\delta_Eu^\mu \left( \theta | E, \ell\right)$ follow geodesics of $\mathcal{V}_{\rm rel}$ :

Considering $u^\mu \left( \theta | E, \ell\right)$ as a function of $E$, holding $\theta$ and $\ell$ arbitrarily fixed, varying $E$ it traverses a space-like directed orbit on $\mathcal{V}_{\rm rel}$. Since $\left|\delta u / \delta E\right| = \Lambda^{-1}$ it is convenient to define an affine parameter $\lambda$ satisfying
\begin{equation}\label{eq: dellambdE}
 \frac{\delta E}{\delta \lambda} = \Lambda \left( E,\ell \right) \, ,
\end{equation}
so that $\left|\delta u / \delta \lambda\right| = 1$, and equation \eref{eq: deludelE} becomes
\begin{equation}\label{eq: deludelam}
 \frac{\delta u^\mu}{\delta \lambda} = \frac{E u^\mu - m v_o^\mu}{\Lambda \left( E,\ell \right)} \, .
\end{equation}
Further variation of \eref{eq: deludelam} yields
\begin{equation}\label{eq: deldeludelam}
 \frac{\delta}{\delta \lambda} \left( \frac{\delta u^\mu}{\delta \lambda} \right) = \frac{\delta}{\delta \lambda} \left( \frac{E u^\mu - mv_o^\mu}{\Lambda \left( E,\ell \right)} \right) = u^\mu \, ,
\end{equation}
which is the geodesic equation on $\mathcal{V}_{\rm rel}$, in accordance with \eref{eq: geodeq}. $u^\mu \left( \theta | E, \ell\right)$ as a function of $E$ is therefore a geodesic on $\mathcal{V}_{\rm rel}$, and may be cast into the generic geodesic form \eref{eq: geogen},
\begin{equation}\label{eq: ugeogen}
 u^\mu\left( \theta | E, \ell\right) = \cosh\lambda\left(E| \ell\right) a^\mu\left( \theta | \ell\right) + \sinh\lambda\left(E| \ell\right) b^\mu\left( \theta | \ell\right) \, .
\end{equation}
The explicit relationship between the vectors $a^\mu(\theta | \ell)$ and $b^\mu(\theta | \ell)$ with $v_o^\mu (\theta |\ell)$ and $n_o^\mu (\theta |\ell)$, as well as the energy dependence of the parameter $\lambda$, depends on the ratio $|\kappa| / \ell$, and is elaborated in the following section for each case separately.

The symmetry group of ${\cal V}_{\rm rel}$, as a 3-D hyperboloid of revolution embedded in a 4-D Euclidian space, is the orthochronous Lorentz group. The geodesics are the integral curves of certain generators of the group. A generic energy variation of the hodograph, now in the form
\begin{equation}\label{eq: delugeogen}
 \delta_E u^\mu\left( \theta | E, \ell\right) = \frac{\delta E}{\Lambda} \sinh\lambda\left(E| \ell\right) a^\mu\left( \theta | \ell\right) + \frac{\delta E}{\Lambda} \cosh\lambda\left(E| \ell\right) b^\mu\left( \theta | \ell\right)
\end{equation}
is recognized (see \ref{sec: Lorgeo}) as the action of a Lorentz transformation on $u^\mu$ via the operator
\begin{equation} \label{eq: Lorop}
 \mathcal{O} = \frac{\delta E}{\Lambda} \left[(b \cdot u) a^\mu - (a \cdot u) b^\mu \right] \frac{\partial}{\partial u^\mu} \, .
\end{equation}

The Newtonian Hamilton symmetry is nothing but the Galilei transformation for velocities. It was referred to so far as transformation between two energy states, but, being uniform and therefore global, it may also be regarded as an active transformation of the same physical system from one inertial frame to another. On the RVS, on the other hand, the Lorentz operator \eref{eq: Lorop} is local, $\theta$-dependent through $a^\mu$ and $b^\mu$, and cannot induce active Lorentz transformations to other inertial frame. The relativistic hodograph method is therefore confined to the centre-of-force rest-frame.

\vskip20pt

\section{Geometrical interpretations of the hodographs and symmetry considerations}\label{sec: geoint}

\begin{figure}[h]
  \centering
  \includegraphics[width=6.5cm]{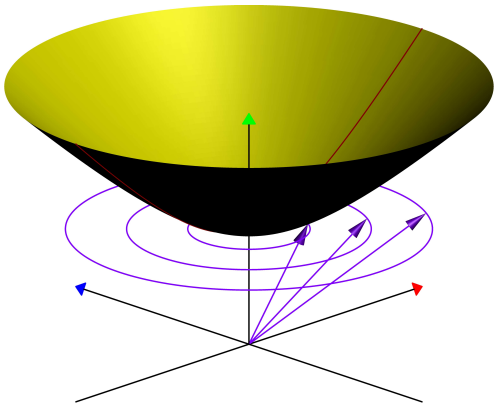}
  \begin{minipage}[b]{255pt}
  \caption{\label{fig:22} \textbf{The unit-velocity hyperboloid ${\cal V}_{\rm rel}$ embedded in the pseudo-Euclidean $E^{(1,3)}$ space.} The axis vector $v_o^\mu$ and the base circle $C_o\left(\ell\right)$ that it traverses, relative to ${\cal V}_{\rm rel}$, for the 3 possible cases :
  (i) left arrow -- $v_o^\mu$ time-like ($\ell > |\kappa|$), if continued the vector punches the hyperboloid into its interior;
  (ii) middle arrow -- $v_o^\mu$ light-like ($\ell = |\kappa|$), if continued the vector approaches the hyperboloid asymptotically;
  (iii) right arrow -- $v_o^\mu$ space-like ($\ell < |\kappa|$), if continued the vector recedes from the hyperboloid.
   The $E^{(1,3)}$-axes with red, blue and green arrowheads correspond, respectively, to $w_x$, $w_y$ and $w^0$ (the $w_z$ axis suppressed).}
  \end{minipage}
\end{figure}

\begin{figure}
  \centering
  \subfloat[Hodograph for bound state. \label{fig:33a}]{\includegraphics[width=4.5cm]{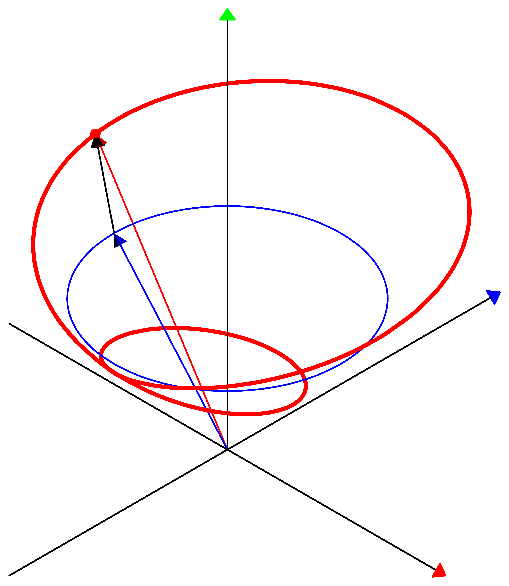}} \quad
  \subfloat[Hodograph for attraction, unbound state. \label{fig:33b}]{\includegraphics[width=4.5cm]{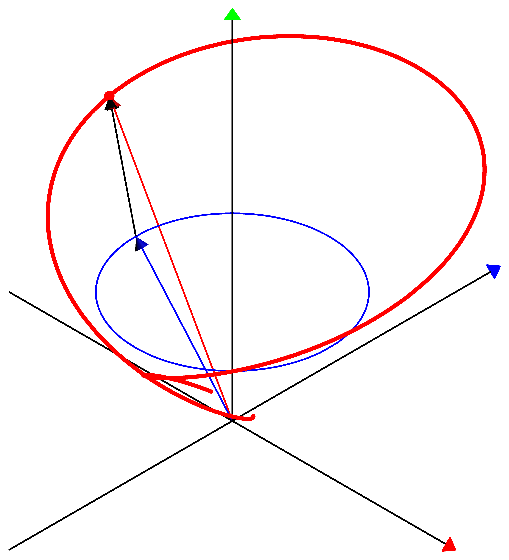}} \quad
  \subfloat[Hodograph for repulsion. \label{fig:33c}]{\includegraphics[width=4.5cm]{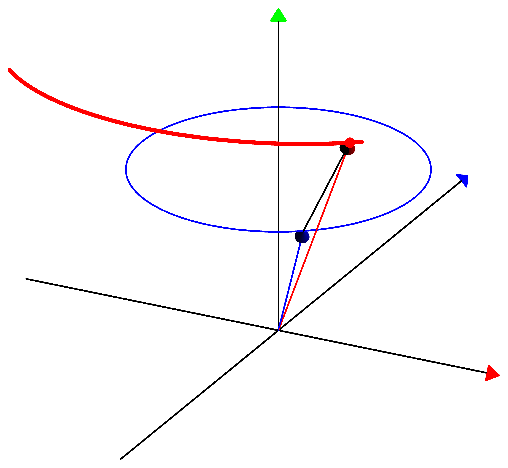}}
  \caption{\label{fig:33} Hodographs for $v_o^\mu$ time-like ($\ell > |\kappa|$) as the superposition $u^\mu (\textrm{red}) = q^\mu (\textrm{blue}) + B^\mu (\textrm{black})$. The $q^\mu$-circle is in the interior of the $\mathcal{V}_{\rm rel}$-hyperboloid.}
\end{figure}

\begin{figure}
  \centering
  \subfloat[Hodograph for bound, unstable state. \label{fig:44a}]{\includegraphics[width=4.5cm]{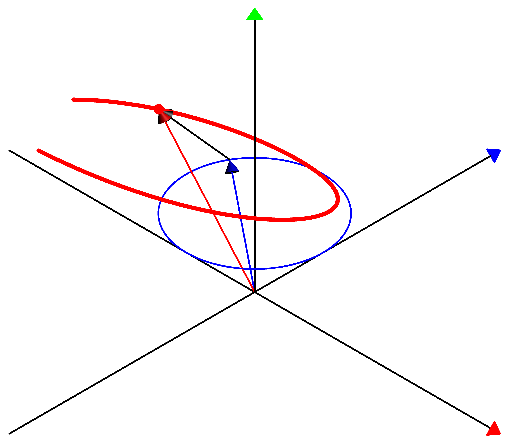}} \quad
  \subfloat[Hodograph for attraction, unbound state. \label{fig:44b}]{\includegraphics[width=4.5cm]{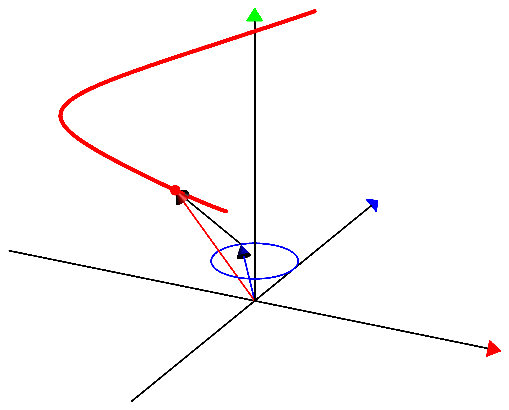}} \quad
  \subfloat[Hodograph for repulsion. \label{fig:44c}]{\includegraphics[width=4.5cm]{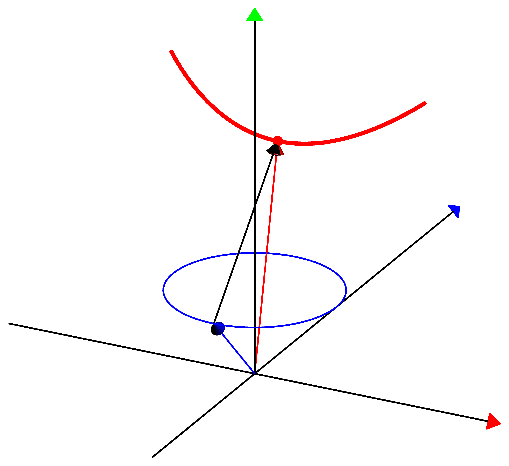}}
  \caption{\label{fig:44} Hodographs for $v_o^\mu$ light-like ($\ell = |\kappa|$) as the superposition $u^\mu (\textrm{red}) = q^\mu (\textrm{blue}) + B^\mu (\textrm{black})$. The $q^\mu$-circle is external to the $\mathcal{V}_{\rm rel}$-hyperboloid.}
\end{figure}

\begin{figure}
  \centering
  \subfloat[Hodograph for bound, unstable state. \label{fig:55a}]{\includegraphics[width=4.5cm]{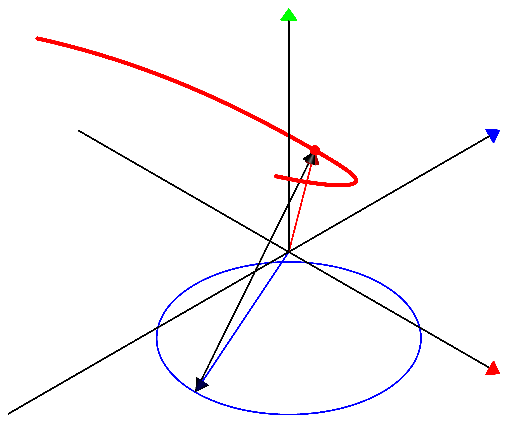}} \quad
  \subfloat[Hodograph for attraction, unbound state. \label{fig:55b}]{\includegraphics[width=4.5cm]{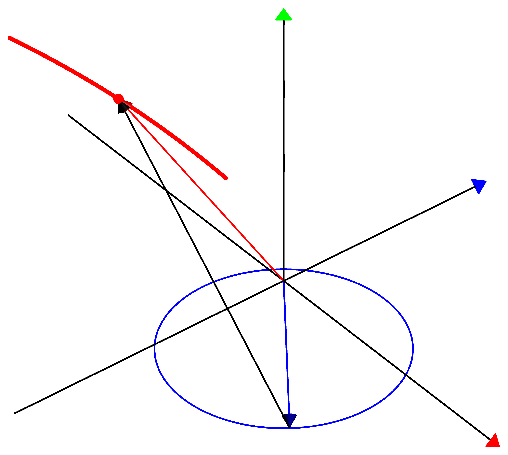}} \quad
  \subfloat[Hodograph for repulsion. \label{fig:55c}]{\includegraphics[width=4.5cm]{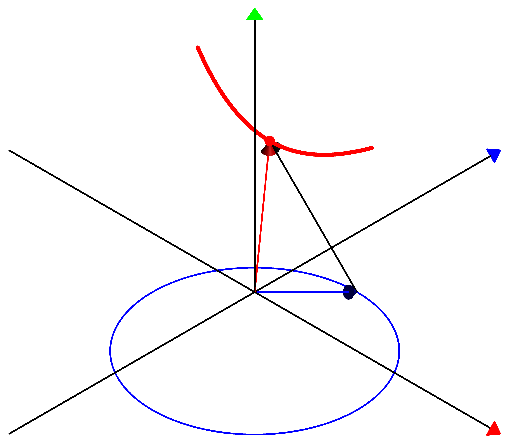}}
  \caption{\label{fig:55} Hodographs for $v_o^\mu$ space-like ($\ell < |\kappa|$) as the superposition $u^\mu (\textrm{red}) = q^\mu (\textrm{blue}) + B^\mu (\textrm{black})$. The $q^\mu$-circle lies below the $w^0 = 0$-plane of $E^{(1,3)}$, the vector $q^\mu$ pointing downwards.}
\end{figure}

In the previous sections we have shown that the relativistic Hamilton symmetry manifests via two aspects :

\begin{enumerate}
 \item {At each instance, the velocity 4-vector $u^\mu$ is the superposition \eref{eq: genhod} of two vectors of constant magnitude and relative angle. One of these vectors ($q^\mu$) is associated with the base circle $C_o(\ell)$, while the other ($B^\mu$) displaces the $q^\mu$-circle to form $u^\mu$.}
 \item {Energy variations of the hodographs are along geodesics of the RVS, generated by Lorentz transformations.}
\end{enumerate}
It will now be shown how these aspects manifest in the relativistic hodographs. The nature of the hodograph depends on the ratio $|\kappa| / \ell$. This is directly related with the magnitude of the axis vector \eref{eq: vodef} :
\begin{equation}\label{eq: vomag}
{v_o}^2 = \frac{\kappa^2}{\ell^2} - 1 \quad \begin{array}{*{20}{c}}
 \nearrow \\
 \to \\
 \searrow
\end{array} \quad \begin{array}{*{20}{c}}
{\ell > |\kappa|}&{\Leftrightarrow}&{v_o^\mu \, {\textrm {time-like}}}\\
{\ell = |\kappa|}&{\Leftrightarrow}&{v_o^\mu \, {\textrm {light-like}}}\\
{\ell < |\kappa|}&{\Leftrightarrow}&{v_o^\mu \, {\textrm {space-like}}}
\end{array}
\end{equation}
$|\kappa| / \ell$ is the magnitude of the spatial velocity on the base circle $C_o(\ell) = \left\{v_o^\mu(\theta)\right\}$, whose projection on the $w_x$-$w_y$ ($w^0 = 0$)-plane in $E^{(1,3)}$ is the minimal energy circle in the Newtonian limit $\left\{\vec v_o(\theta)\right\}$ (the same notation $C_o(\ell)$ is used for both circles for convenience). Only for $\ell >|\kappa|$ are the velocities on $C_o(\ell)$ subluminal. Therefore, the relativistic requirement that particles' velocities cannot reach the velocity of light necessarily implies that the relativistic solution can have a non-relativistic limit only for $v_o^\mu$ time-like ($\ell > |\kappa|$). Geometrically, the vector $v_o^\mu$ then points towards the interior of $\mathcal{V}_{\rm rel}$ (see \Fref{fig:22}) so that, if continued to form $q^\mu$, it punches through the hyperboloid into its interior and draws on it a horizontal circle which is the hodograph
\begin{equation}\label{eq: uodef}
 u_o^\mu(\theta|\ell) \equiv \frac{1}{\beta} v_o^\mu(\theta|\ell) = \left( \frac{1}{\beta}, - \frac{\kappa}{\beta \ell} \hat\theta \right) \quad , \quad \beta = \sqrt {1 - \frac{\kappa^2}{\ell^2}} \, .
\end{equation}
Then $B^\mu$ superposes upon $q^\mu$ complementing it from the interior of the hyperboloid to form the hodograph $u^\mu$ according to \eref{eq: genhod} (see \Fref{fig:33})\footnote{\Fref{fig:33}, \ref{fig:44} \& \ref{fig:55} illustrate the superposition $u^\mu = q^\mu + B^\mu$ \eref{eq: genhod} for the various cases. A common colour code is used: the vector $q^\mu$ and the circle it draws in blue, the vector $B^\mu$ in black, and the velocity vector $u^\mu$ and the hodograph in red. These colours may not be seen in printed versions, and the reader is advised to look at the online or PDF versions.}.

Otherwise, for $\ell \le |\kappa|$, $v_o^\mu$ points outside of the hyperboloid (see \Fref{fig:22}) : a light-like $v_o^\mu$ (for $\ell = |\kappa|$), if continued, approaches the hyperboloid asymptotically, while a space-like $v_o^\mu$ (for $\ell < |\kappa|$), if continued, recedes from the hyperboloid. In either case, $q^\mu$ draws a circle in the exterior of the hyperboloid and from there it is complemented by $B^\mu$, in these cases pointing inwardly (see \Fref{fig:44} \& \ref{fig:55}). Geometrically this is the main difference that eventually marks the distinction between the hodographs and trajectories in the different cases.

\subsection{Hodographs with Newtonian limit ($\ell > |\kappa|$)} \label{sec: geoellarge}

We start with the hodographs with Newtonian limit. These were found \cite{Relhod} as
\begin{equation}\label{eq: usol}
\begin{eqalign}
 {\vec u &=  B_o \sin \left( \beta \theta - \varphi \right) \hat r + \left[ - \frac{\kappa E}{m \ell \beta^2} + \frac{B_o}{\beta} \cos \left( \beta \theta - \varphi \right) \right] \hat \theta \, , \\
 u^0 &= \frac{E}{m} - \frac{\kappa}{\ell} u_\theta = \frac{E}{\beta^2 m} - \frac{\kappa B_o}{\beta\ell} \cos \left( \beta \theta - \varphi \right)}
\end{eqalign}
\end{equation}
with
\begin{equation}\label{eq: betaBo}
 B_o = \sqrt {\frac{E^2}{\beta^2 m^2} - 1} = \frac{\Lambda \left( E, \ell \right)}{\beta m}
\end{equation}
and $\varphi$ is an arbitrary constant shift angle.

For a given value of the angular momentum $\ell$ the hodograph $u^\mu = u_o^\mu(\theta|\ell)$ \eref{eq: uodef} corresponds to the state of minimal energy $E = \beta m$ with $B_ o= 0$, and spatial circular motion. For increased energy $E > \beta m$, the solution \eref{eq: usol} may be written as
\begin{equation}\label{eq: usol2}
 u^\mu(\theta|E,\ell) = \frac{E}{\beta m} u_o^\mu + B_o n_1^\mu = \sqrt{1 + {B_o}^2} u_o^\mu + B_o n_1^\mu \, .
 \end{equation}
with the vector
\begin{equation}\label{eq: n1def}
 n_1^\mu(\theta|\ell) \equiv \left( - \frac{\kappa}{\beta \ell} \cos(\beta \theta - \varphi), \sin(\beta \theta - \varphi) \hat r + \frac{1}{\beta} \cos(\beta \theta - \varphi) \hat \theta \right) \, .
\end{equation}
$n_1^\mu$ is a space-like unit 4-vector, tangent to $\mathcal{V}_{\rm rel}$ at $u_o^\mu$, satisfying $n_1 \cdot n_1 = 1 \, , \, u_o \cdot n_1 = 0$.

The representation \eref{eq: usol2} is in accord with the generic hodograph form \eref{eq: genhod}, with $q^\mu = \left(E/m \beta\right) u_o^\mu$ and $n_o^\mu = n_1^\mu / m \beta$ : $q^\mu$ draws a horizontal circle within the $\mathcal{V}_{\rm rel}$-hyperboloid, and the vector $B^\mu = B_o n_1^\mu$ superposes upon the uniformly rotating vector $q^\mu$ to form the velocity 4-vector $u^\mu$.

The vector $n_1^\mu$ is not constant, but rotates in $E^{(1,3)}$. Together with the unit vector
\begin{equation} \label{eq: n2def}
 n_2^\mu \left(\theta|\ell\right) = \left( \frac{\kappa}{\beta \ell} \sin \left( \beta \theta - \varphi \right), \cos \left( \beta \theta - \varphi \right) \hat r - \frac{1}{\beta} \sin \left( \beta \theta - \varphi \right) \hat \theta \right)
\end{equation}
(which may be obtained from $n_1^\mu\left(\theta|\ell\right)$ by substituting $\varphi \rightarrow \varphi - \pi/2$ in \eref{eq: n1def}) and $n_3^\mu = \left(0,\hat z\right)$, the triad $\left\{n_i^\mu \, , \, i=1,2,3\right\}$ spans ${\cal T}_{u_o} \left( {\cal V}_{\rm rel} \right)$, the hyperplane tangent to ${\cal V}_{\rm rel}$ at $u_o^\mu$. The equations of motion of these vectors are
\begin{equation}\label{eq: n1n2eq}
\begin{eqalign}
 { \frac{Dn_1^\mu}{d\theta} &= \frac{dn_1^\mu}{d\theta} - \left( n_1 \cdot \frac{d u_o}{d\theta} \right) u_o^\mu = - \frac{\kappa^2}{\beta \ell^2} n_2^\mu \\
 \frac{Dn_2^\mu}{d\theta} &= \frac{dn_2^\mu}{d\theta} - \left( n_2 \cdot \frac{d u_o}{d\theta} \right) u_o^\mu = \frac{\kappa^2}{\beta \ell^2} n_1^\mu }
\end{eqalign}
\end{equation}
$n_3^\mu$ obviously remains constant. The lhs of these equations are the tangential derivatives of $n_1^\mu$ and $n_2^\mu$ on the tangent hyperplane ${\cal T}_{u_o} \left( {\cal V}_{\rm rel} \right)$, indicating that these vectors rotate uniformly in the hyperplane. The rotation may be explicitly demonstrated by introducing the unit vectors
\begin{equation}\label{eq: e1e2def}
\begin{eqalign}
 {\fl \hskip20pt e_1^\mu \left(\theta|\ell\right) \equiv \left( - \frac{\kappa}{\beta \ell} \cos \left( \beta^{-1} \theta - \varphi \right), \sin \left( \beta^{-1} \theta - \varphi \right) \hat r + \beta^{-1} \cos \left( \beta^{-1} \theta - \varphi \right) \hat \theta \right) \\
 \fl \hskip20pt e_2^\mu \left(\theta|\ell\right) \equiv \left( \frac{\kappa}{\beta \ell} \sin \left( \beta^{-1} \theta - \varphi \right), \cos \left( \beta^{-1} \theta - \varphi \right) \hat r - \beta^{-1} \sin \left( \beta^{-1} \theta - \varphi \right) \hat \theta \right) }
\end{eqalign}
\end{equation}
which together with $e_3^\mu = n_3^\mu$ form a Fermi-Walker orthonormal triad $\left\{e_i^\mu \, , \, i=1,2,3\right\}$ that also spans ${\mathcal T}_{u_o} \left( {\cal V}_{\rm rel} \right)$. It is then straight-forward to show that
\begin{equation}\label{eq: n1n2-e1e2}
\begin{eqalign}
 { n_1^\mu = \cos \left( \frac{\kappa^2}{\beta \ell^2} \theta \right) e_1^\mu - \sin \left( \frac{\kappa^2}{\beta \ell^2} \theta \right) e_2^\mu \\
 n_2^\mu = \sin \left( \frac{\kappa^2}{\beta \ell^2} \theta \right) e_1^\mu + \cos \left( \frac{\kappa^2}{\beta \ell^2} \theta \right) e_2^\mu }
\end{eqalign}
\end{equation}

The representation \eref{eq: usol2} is also in accord with the generic geodesic form \eref{eq: ugeogen}, identifying the vectors $a^\mu = u_o^\mu$ and $b^\mu = n_1^\mu$ and the affine parameter $\lambda (E |\ell)$ satisfying
\begin{equation}\label{eq: lambdaE}
 \cosh\lambda (E|\ell) = \frac{E}{\beta m} = \sqrt{1 + {B_o}^2} \quad , \quad \sinh\lambda (E|\ell) = B_o = \frac{\Lambda}{\beta m}
\end{equation}
$\lambda = 0$ corresponds to the state of minimum energy. Then, for each $\theta$, a Lorentz transformation in the direction of $n_1^\mu(\theta|\ell)$ takes the hodograph from the minimum energy state $u_o^\mu(\theta|\ell)$ to the actual hodograph $u^\mu(\theta|E,\ell)$ along the geodesic whose tangent is
\begin{equation}\label{eq: delulam}
 \frac{\delta u^\mu}{\delta \lambda} = \frac{E}{\beta m} n_1^\mu - \frac{\Lambda}{\beta m} u_o^\mu = \sqrt{1 + {B_o}^2} n_1^\mu - B_o u_o^\mu \, .
 \end{equation}
\eref{eq: delulam} is the explicit expression for \eref{eq: deludelam}.

\subsection{Hodographs for $\ell = |\kappa|$} \label{sec: geoelleqk}

The hodograph solution for $\kappa = \pm \ell$ was found \cite{Relhod} as
\begin{equation}\label{eq: ukeql}
\begin{eqalign}
 { \vec u = \pm \frac{E}{m} \left( \theta - \theta_o \right) \hat r \mp \left[ \frac{E}{2m} \left( \theta - \theta_o \right)^2 - \frac{E^2 - m^2}{2mE} \right] \hat \theta \\
 u^0 = \frac{E}{m} \mp u_\theta = \frac{E}{2m} \left(\theta - \theta_o \right)^2 + \frac{E^2 + m^2}{2mE} }
\end{eqalign}
\end{equation}
with $\theta_o$ an arbitrary shift angle. It may be combined in the form \eref{eq: genhod} as
\begin{equation}\label{eq: ukeql2}
 u^\mu (\theta | E) = \frac{m}{2E} v_o^\mu + \frac{E}{2m} n^\mu
\end{equation}
with $n^\mu$ defined by
\begin{equation}\label{eq: nokeql}
 n^\mu (\theta) = \left( 1 + (\theta - \theta_o)^2, \pm 2(\theta - \theta_o) \hat r \pm \left[ 1 - (\theta - \theta_o)^2 \right] \hat\theta \right)
\end{equation}

In the present case $\Lambda(E) = E$, so that by \eref{eq: dellambdE} $\lambda(E) = \ln(E/m)$. The energy may get any value $0 < E < \infty$ for the $\lambda$-range $-\infty < \lambda < \infty$. Both vectors $v_o^\mu$ and $n^\mu$ are light-like, satisfying
\begin{equation}\label{eq: novokeql}
 v_o \cdot v_o = n \cdot n = 0 \quad , \quad n \cdot v_o = -2
\end{equation}
Then, with the vectors
\begin{equation}\label{eq: ableqk}
\begin{eqalign}
 { a^\mu = \frac{1}{2} (n^\mu + v_o^\mu) = \left( \frac{1}{2}(\theta - \theta_o)^2 + 1, \pm (\theta - \theta_o) \hat r \mp \frac{1}{2}(\theta - \theta_o)^2 \hat\theta \right) \\
  b^\mu = \frac{1}{2} (n^\mu - v_o^\mu) = \left( \frac{1}{2} (\theta - \theta_o)^2, \pm (\theta - \theta_o) \hat r \pm \left[ 1 - \frac{1}{2}(\theta - \theta_o)^2 \right] \hat\theta \right) }
\end{eqalign}
 \end{equation}
the generic geodesic form \eref{eq: ugeogen} is recovered with
\begin{equation}\label{eq: lambdaEleqk}
 \cosh\lambda(E) = \frac{E^2 + m^2}{2mE} \quad , \quad \sinh\lambda(E) = \frac{E^2 - m^2}{2mE}
\end{equation}
For $\lambda = 0$, or $E = m$, the base hodograph is $u^\mu = a^\mu(\theta|\ell)$, and all other hodographs may be obtained geodesically by Lorentz transformations along $b^\mu$ with the unit-velocity variation \eref{eq: deludelam}
\begin{equation}\label{eq: deludelam3}
 \frac{\delta u^\mu}{\delta \lambda} = -\frac{m}{2E} v_o^\mu + \frac{E}{2m} n^\mu \, .
\end{equation}
There is no circular hodograph because $v_o^\mu$ points outside of the ${\cal V}_{\rm rel}$-hyperboloid and there are no stable states.

\subsection{Hodographs for $\ell < |\kappa|$} \label{sec: geoellsmall}

In these cases the axis of rotation $v_o^\mu$ is space-like, pointing outside of the hyperboloid $\mathcal{V}_{\rm rel}$. The hodograph solution is \cite{Relhod}
\begin{equation}\label{eq: uklarge}
\begin{eqalign}
 { \vec u = -\epsilon A_o \sinh \left[\bar\beta \left( \theta - \theta_o \right)\right] \hat r + \left\{ \frac{\kappa E}{m\ell \bar\beta^2} - \frac{\epsilon A_o}{\bar\beta} \cosh \left[\bar\beta \left( \theta - \theta_o \right)\right] \right\} \hat\theta \\
  u^0 = - \frac{E}{\bar\beta^2 m} + \frac{\left| \kappa \right| A_o}{\bar\beta \ell} \cosh \left[\bar\beta \left( \theta - \theta_o \right)\right] }
\end{eqalign}
\end{equation}
with
\begin{equation}\label{eq: betabarAo}
 \bar\beta = \sqrt {\frac{\kappa^2}{\ell^2} - 1} \quad , \quad A_o = \sqrt {\frac{E^2}{\bar\beta^2 m^2} + 1} = \frac{\Lambda}{\bar\beta m} \, ,
 \end{equation}
$\epsilon = \rm{sign}(\kappa)$ and $\theta_o$ an arbitrary constant shift angle. It coincides with the generic hodograph representation \eref{eq: genhod} as
\begin{equation}\label{eq: uklarge2}
 u^\mu (\theta | E,\ell) = - \frac{E}{\bar\beta^2 m} v_o^\mu + A_o n^\mu
\end{equation}
with
\begin{equation}\label{eq: wnklarge}
 \fl \hskip15pt n^\mu (\theta | \ell) = \left( \frac{\left| \kappa \right|}{\bar\beta \ell} \cosh \left[\bar\beta (\theta - \theta_o)\right], \epsilon \sinh \left[\bar\beta (\theta - \theta_o)\right] \hat r - \frac{\epsilon}{\bar\beta} \cosh \left[\bar\beta (\theta - \theta_o)\right] \hat\theta \right)
\end{equation}
The roles of $v_o^\mu$ and $n^\mu$ are now interchanged, since $v_o^\mu$ is space-like and $n^\mu$ is a time-like unit 4-vector. The negative coefficient of $v_o^\mu$ turns it up-side down, so that $q^\mu$ points downwards rather than upwards.

The affine parameter $\lambda (E,\ell)$ is found from \eref{eq: dellambdE},
\begin{equation}\label{eq: lambdaEllek}
 \cosh\lambda (E,\ell) = \frac{\sqrt {E^2 + \bar\beta^2 m^2}}{\bar\beta m} = A_o \quad , \quad \sinh\lambda (E,\ell) = \frac{E}{\bar\beta m} \, ,
\end{equation}
and the generic geodesic form \eref{eq: ugeogen} is recovered, identifying $a^\mu = n^\mu$ and $b^\mu = - \bar\beta^{-1} v_o^\mu$. For $\lambda = 0$, or $E = 0$, the base hodograph is $u^\mu = a^\mu(\theta|\ell)$, and all other hodographs may be obtained geodesically by Lorentz transformations along $b^\mu$ with the unit-velocity variation \eref{eq: deludelam}
\begin{equation}\label{eq: deludelam2}
 \frac{\delta u^\mu}{\delta \lambda} = \frac{E}{\bar\beta m} n^\mu - \frac{A_o}{\bar\beta} v_o^\mu \, .
\end{equation}
As in the former case there is no circular hodograph, because $v_o^\mu$ points outside of the ${\cal V}_{\rm rel}$-hyperboloid and there are no stable states.

\vskip20pt

\section{Relativistic Hamilton vector} \label{sec: RelHamvec}

Finally, having demonstrated the action of Hamilton symmetry in relativistic velocity space, it is in place to identify the major actor, the {\it relativistic Hamilton vector}. Following the Newtonian Hamilton symmetry and the previous sections, it is natural to expect this vector to be associated with the geodesic translation of the hodographs in velocity space.

\subsection{Identification of the Hamilton vector} \label{sec: idhamvec}

We start as above with the cases with Newtonian limit, $\ell > \left|\kappa\right|$. The hodograph is decomposed \eref{eq: usol2} as $u^\mu(\theta) = q^\mu + B^\mu$, where $q^\mu = \left(E/\beta^2 m\right) v_o^\mu$ draws the straight-forward extension of the base circle $C_o(\ell)$, and $B^\mu = B_o n_1^\mu$ is directed along the energy-dependent geodesic.

There is a good geometrical reason why the base, $q$-drawn, circle is not an hodograph but rather lies within the velocity hyperboloid ${\cal V}_{\rm rel}$ : Because of the curvature of the hyperboloid, the vector $B^\mu$, if it started at $u_o^\mu$, would point outwards of the hyperboloid. $q^\mu$ punches the hyperboloid at $u_o^\mu$ and enters into it in just the right amount so that the combination $q^\mu + B^\mu$ is an hodograph on ${\cal V}_{\rm rel}$. It is therefore appropriate to regard
\begin{equation}\label{eq: relHvec}
 \fl \hskip20pt B^\mu = B_o n_1^\mu = \left( - \frac{\kappa B_o}{\beta \ell} \cos \left( \beta \theta - \varphi \right), B_o \sin \left( \beta \theta - \varphi \right) \hat r + \frac{B_o}{\beta} \cos \left( \beta \theta - \varphi \right) \hat \theta \right)
 \end{equation}
as the {\it relativistic Hamilton vector}.

In the Newtonian limit $B_o n_1^\mu \to \left(0,B_o \hat\varphi\right)$, with the Newtonian limit of $B_o$ given by \eref{eq: BoNewt} and $u^\mu \to \left(1, \vec v_o(\theta) + B_o \hat\varphi\right)$. $\vec B_o = B_o\hat\varphi$ is identified, according to \eref{eq: hodoN}, as the classical Hamilton vector, so the appropriate limit is verified. Also, in the Newtonian hodograph \eref{eq: hodoN} the energy dependence is only through $B_o$ which may be identified as the affine parameter $\lambda (E',\ell) = B_o$ for the straight line drawn by $\vec B_o$. The unit velocity variation, the Newtonian limit of \eref{eq: deludelam}, is then $\delta \vec v / \delta \lambda = \hat \varphi$. Variation with $\lambda$ implies variation of the energy without changing the angular momentum, again recognized as the Hamilton symmetry.

Things are less clear-cut for the exclusive relativistic cases $\ell \le \left|\kappa\right|$. Here there is no circular base hodograph like $u_o^\mu$. The $q^\mu$-drawn circles, with $q^\mu = - \left(E/\bar\beta^2 m\right) v_o^\mu$ or $q^\mu = (m/2E) v_o^\mu$ for $\ell < |\kappa|$ or $\ell = |\kappa|$, respectively, lie outside the ${\cal V}_{\rm rel}$ hyperboloid. The completion vectors, $B^\mu = (E/2m) n^\mu$ for $\ell = |\kappa|$ \eref{eq: ukeql2} and $B^\mu = A_o n^\mu$ for $\ell < |\kappa|$ \eref{eq: uklarge2}, reside outside the hyperboloid and do not generate the energy-dependent geodesics. The $B^\mu$ vectors could be formally defined, based on the decomposition $u^\mu = q^\mu + B^\mu$, as Hamilton vectors, but without a clear significance of this definition. Since the identification of an Hamilton vector much relies on the Newtonian limit, the question of identifying an Hamilton vector for the exclusive relativistic cases is left open.

\subsection{Relation with the Laplace-Runge-Lenz symmetry} \label{sec: LRLsym}

The Hamilton symmetry is an internal symmetry of KC systems, over and beyond the central symmetry of the $1/r$ potential. As is well known, the $1/r$ interaction allows Newtonian KC systems to enjoy a very special and simple form of the Laplace-Runge-Lenz (LRL) symmetry, which is also directly associated with the spatial trajectories being conic sections \cite{McIntosh71,GoldsteinLRL,Goldstein.etal2000,KhachidzeKhelash08}. Both symmetries (LRL and Hamilton) are indeed closely related :

The (constant) LRL vector for Newtonian KC systems,
 \begin{equation} \label{eq: NewtLRL}
 \vec K_o = \vec p \times \vec\ell + m\kappa \hat r \, ,
 \end{equation}
points along the major axis of the (conic sections) spatial trajectories. Expressing $\vec B_o$ in the Newtonian hodograph solution \eref{eq: hodoN} in terms of the phase-space variables,
 \begin{equation} \label{eq: NewtHvec}
 \vec B_o = \frac{\vec p}{m} +\frac{\kappa}{\ell^2} \vec\ell \times \hat r \, ,
 \end{equation}
the relation between the two vectors is immediately identified as $\vec K_o = - m\vec \ell \times \vec B_o$ \cite{Munoz2003}. Both vectors, perpendicular to each other and to $\vec \ell$, are equivalently associated with the shape of the conic sections.

The relation between the two symmetries extends into the relativistic realm. The LRL symmetry in relativistic Coulomb systems was studied to some extent in recent years \cite{Yoshida1988,Stahlhofen2005a,LRLgen}, limited to the cases with Newtonian limit (here also it is not clear how to define an LRL vector for the exclusive relativistic cases). A relativistic LRL vector, pointing along the major axis of the (rotating conic sections) spatial trajectories, was identified,
 \begin{equation} \label{eq: rotrelLRL}
 \vec K = \beta \vec p \times \vec \ell + \left( \kappa E - \frac{\beta\Omega \ell^2}{r} \right) \hat r
 \end{equation}
with $\Omega = 1 - \beta$. This vector has constant magnitude $\left|\vec K\right| = \ell\Lambda$, but it rotates with the major axis of the spatial trajectory. The rotation may be compensated using the rotation operator
 \begin{equation} \label{eq: rotop}
 {\mathcal U}(\psi) \equiv \cos \psi + \sin \psi \hat \ell \times
 \end{equation}
for a rotation angle $\psi$ in the plane of motion, and a constant relativistic LRL vector is identified
 \begin{equation} \label{eq: relLRL}
 \vec K_o = {\mathcal U} (- \Omega \theta) \vec K = {\mathcal U} (- \Omega \theta) \left[ \beta \vec p \times \vec \ell + \left( \kappa E - \frac{\beta\Omega \ell^2}{r} \right) \hat r \right]
 \end{equation}
directed along the major axis at some particular point in the evolution of the system.

Since the directional freedom of the Hamilton vector \eref{eq: relHvec} is in the constant unit vector $\hat \varphi$, we expect this vector to be related with the relativistic LRL vector. It is easily verified that
\begin{equation}\label{eq: Omphi}
\begin{eqalign}
 { \sin(\beta \theta - \varphi) = \hat r \cdot \left[ {\mathcal U} \left( \Omega \theta \right) \hat\varphi \right] \\
 \cos(\beta \theta - \varphi) = \hat \theta \cdot \left[ {\cal U} \left( \Omega \theta \right) \hat\varphi \right] }
\end{eqalign}
\end{equation}
With $\vec B_o = B_o \hat\varphi$, substituting \eref{eq: Omphi} in \eref{eq: usol} yields, after some algebra, the relation
\begin{equation}\label{eq: UBo}
 {\mathcal U} \left( \Omega \theta \right) \vec B_o = u_r \hat r + \left( \beta u_\theta + \frac{\kappa E}{\beta m\ell} \right) \hat \theta = \vec u + \left( \frac{\kappa E}{\beta m\ell} - \Omega u_\theta \right) \hat \theta \, ,
\end{equation}
and using the relations $\vec p = m \vec u$, $u_\theta = \ell / m r$, yields
 \begin{equation} \label{eq: relBo}
 m {\mathcal U} (\Omega \theta) \vec B_o = \vec p + \left( \frac{\kappa E}{\beta\ell} - \frac{\Omega \ell}{r} \right) \hat \theta = \frac{\vec\ell \times\vec K}{\beta\ell^2}
 \end{equation}
Then it is straight-forward to verify the relation
 \begin{equation} \label{eq: relKo}
 \vec K_o = - \beta m\vec \ell \times \vec B_o
 \end{equation}
which also verifies the constancy of $\vec K_o$.

The relation between the symmetries extends into the full space-time picture. The relativistic LRL 4-vector is fully space-like in the centre-of-mass (CM) reference frame (or centre-of-force in the present case), so the constant LRL 4-vector is $K_o^\mu = \left(0,\vec K_o\right)$ and the rotating 4-vector is $K^\mu = \left(0,\vec K\right)$. Then the relation between the generating 4-vectors is (the $z$-axis suppressed)
 \begin{equation} \label{eq: **}
{B^\mu } = \frac{1}{\beta^2 m \ell}
 \left( {\begin{array}{*{20}{c}}
 {1}&{- \frac{\kappa}{\ell}\cos\theta}&{- \frac{\kappa}{\ell}\sin\theta} \\
 {\frac{\kappa}{\ell}\sin \theta }&{- \Omega\sin\theta \cos\theta }&{-\beta - \Omega\sin^2\theta} \\
{ - \frac{\kappa}{\ell}\cos\theta}&{\beta + \Omega\cos^2\theta}&{\Omega\sin\theta \cos\theta}
\end{array}} \right)
\left( {\begin{array}{*{20}{c}}
0\\
{K_x}\\
{K_y}
\end{array}} \right)
 \end{equation}

The one-to-one correspondence between the generating vectors implies the equivalence of Hamilton and LRL symmetries. The equivalence is essentially mathematical, while physically there is also a significant distinction : The Hamilton symmetry, acting in velocity space, transforms between states of same angular momentum but different energies, while the LRL symmetry, acting in configuration space, transforms between states of same energy but different internal angular momentum. The two symmetries therefore complement each other.

\subsection{Alternative definitions of the Hamilton vector} \label{sec: althamvec}

Unlike the Newtonian limit, the $q$-drawn base circle in the hodograph decomposition $u^\mu(\theta) = q^\mu + B^\mu$ is energy dependent. An alternative decomposition
\begin{equation}\label{eq: usol3}
 \hskip30pt u^\mu(\theta|E,\ell) = u_o^\mu + \left[\left(\frac{E}{\beta m} - 1\right) u_o^\mu + \frac{\Lambda(E,\ell)}{\beta m} n_1^\mu\right]
 \end{equation}
could be considered, which uses the (energy-independent) minimum energy hodograph -- the $u_o^\mu$-drawn circle -- as a base circle, but here the completion vector (the vector in the square brackets) is not in the direction of a geodesic and cannot be regarded as generating the symmetry. Also, it cannot be related in a simple way to the LRL vector, thus excluding its candidacy for the Hamilton vector.

The only other instance known to the author with an attempt to identify and define a relativistic Hamilton vector is in a paper by Mu\~noz and Pavic \cite{MunozPavic2006} which also discusses extending Hamilton's method to relativistic Coulomb systems. Their approach, however, differs from the present paper in some major aspects :

First, their discussion is 3-D, without reference to the relativistic velocity space and its relation with the structure of Minkowski space-time, and with no reference to Hamilton symmetry. Then, while their solutions for $\vec u(\theta)$ are correct, they defined an Hamilton vector $\vec h$ as an {\it ad-hoc} adaptation of the non-relativistic relation \eref{eq: hodoN}, which in our notation is
\begin{equation}\label{eq: MPHvec}
 \vec h = \vec u + \frac{\kappa}{\ell} u^o \hat \theta \, .
\end{equation}
This vector, differing from the spatial part of \eref{eq: relHvec}
\begin{equation}\label{eq: vecB}
 \vec B = u_r \hat r + \left( u_\theta + \frac{\kappa E}{m\ell \beta^2} \right) \hat \theta = \vec u + \frac{\kappa E}{m\ell \beta^2} \hat \theta \, ,
\end{equation}
does not generate, in any case, Hamilton's symmetry, it also cannot be related in a simple way to the LRL vector, and it is not clear if it has any geometrical significance, even for the cases with Newtonian limits. It is the author's belief, therefore, that the vector $B^\mu$ \eref{eq: relHvec} provides a more natural geometrical interpretation of the relativistic hodograph, as discussed above. Nevertheless, Mu\~noz and Pavic examine their results in a wide range of the parameters $\kappa/\ell$ and $E/m$ with many illustrations for $\vec u(\theta)$, which the reader may well benefit from.

\vskip20pt

\section{Concluding remarks} \label{sec: Con}

Hamilton's hodograph method reveals, when applied to Newtonian Kepler/Coulomb (KC) systems, the internal symmetry which translates and connects between different energy states, same angular momentum, in velocity space. The simplicity and elegance of this application prompt naturally an interest in its extension to relativistic Coulomb systems, which is the purpose of the present research. The results seem very satisfactory:  Although the spatial trajectories of relativistic Coulomb systems are much more complicated than the Newtonian ones, the hodograph symmetry maintains its two main features -- general hodographs may be represented as linearly displaced base energy-independent circles, and hodographs corresponding to states with same angular momentum but different energies are connected via translations along geodesics of the velocity space.

The unique relativistic features are that the velocity space is hyperbolic with Lorentz-generated geodesics, and the hodograph displacement is itself rotating, so the whole phenomenon is of rotation superimposed on rotation -- precession on the velocity hyperboloid, reminding in a way the ancient Greek picture of epicyclic planetary trajectories.

Another major feature of the Hamilton symmetry is its equivalence with the Laplace-Runge-Lenz (LRL) symmetry : For Newtonian KC systems the combination of central symmetry with the extra LRL symmetry provides the full solution in configuration space, with the spatial trajectories being conic sections. Similarly, using the central symmetry (angular momentum conservation) to confine the analysis to velocity space, produces the hodograph method with the manifest Hamilton symmetry, again providing the full solution. The equivalence of both symmetries for relativistic Coulomb systems was demonstrated above.

Coulomb systems are the limit of 2-body EM systems when one of the bodies is much more massive than the other. Besides the interest in the relativistic Hamilton symmetry in Coulomb systems \textit{per se}, the present study is also part of an attempt to advance the solution of the relativistic general EM 2-body problem : Unlike the Newtonian case, the relativistic (non-quantum) EM 2-body problem doesn't have, despite all the years, a satisfactory solution. Some simple particular solutions have been found, especially in the 60's and the 70's \cite{Stephas1978}, but since then no real advancement has been marked and a general solution is missing. The success of using the internal symmetries for a complete solution of the Newtonian KC systems and their extension for relativistic Coulomb systems gives rise to the hope that the corresponding extra symmetries may assist in advancing a solution also for more general relativistic EM systems.

Central symmetry is used in the hodograph method to confine the analysis to velocity space, eliminating using the time as an evolution parameter. For relativistic systems, enjoying Lorentz-Poincare symmetry, central symmetry is a manifestation of Wigner's little group, which is the internal reduction of the global rotational symmetry to the centre-of-mass (CM) frame. It has been shown recently \cite{LRLgen} that the (internal) LRL symmetry (now recognized as LRL/Hamilton symmetry) is similarly the reduction of the global Lorentz symmetry to the CM frame, demonstrating another aspect of the relation between Lorentz and Hamilton symmetries.

In the relativistic hodograph method for Coulomb systems the very-massive particle rests at the bottom of the RVS, while the lighter one moves on the hodograph. In a general 2-body system both particles are moving. How may non-instantaneous interactions be handled on velocity space, especially in view of the fact that time doesn't appear in the structure of the velocity space ? This is an intriguing question, to which we hope to come back soon.

\appendix

\section{Geometrical properties of the relativistic velocity space}\label{sec: geoprop}

The relativistic velocity space (RVS) ${\cal V}_{\rm rel}$ is the 3-D unit hyperboloid $\left\{ u^\mu = \left(\sqrt{1 + \vec u\, ^2} , \vec u \right)\right\}$ \eref{eq: Vrel} embedded in a 4-D pseudo-Euclidian space $E^{(1,3)}$ \eref{eq: E13}. For any $u^\mu \in {\cal V}_{\rm rel}$, ${\cal T}_u\left({\cal V}_{\rm rel}\right) \subset E^{(1,3)}$ is the hyperplane tangent to ${\cal V}_{\rm rel}$ at $u^\mu$. The RVS was discussed mainly by Rhodes and Semon \cite{RhodesSemon2004}; see also \cite{LLF5,CriadoAlamo2001,Urbantke1990,Aravind1997}. In the following are listed some features of the RVS that are used in the paper.

\subsection{Line element and metric tensor in ${\cal V}_{\rm rel}$}

On ${\cal V}_{\rm rel}$ $u \cdot u = -1$, therefore $u \cdot du = - u^0 du^0 + \vec u \cdot d\vec u = 0$. Therefore, for an infinitesimal displacement $d\vec u$, the corresponding displacement in $E^{(1,3)}$ tangent to ${\cal V}_{\rm rel}$ is
\begin{equation}\label{eq: Tdis}
 d_{\rm{T}} u^\mu = \left( \frac{\vec u \cdot d\vec u}{\sqrt {1 + \vec u^2 }},d\vec u \right)
\end{equation}
Since $E^{(1,3)}$ is pseudo-Euclidean, the tangent displacement defines the line element in ${\cal V}_{\rm rel}$ by
\begin{equation}\label{eq: dl2}
 d\lambda^2 = \left(d_{\rm{T}} u\right)^2 = \left( d\vec u \right)^2 - \frac{\left( \vec u \cdot d\vec u \right)^2}{1 + \vec u^2}
\end{equation}
with the metric tensor in these coordinates
\begin{equation}\label{eq: gij}
 g_{ij} = \delta_{ij} - \frac{u^i u^j}{1 + \vec u^2}
\end{equation}
for $d\lambda^2 = g_{ij} du^i du^j$. Since the Lorentz group is the symmetry group of ${\cal V}_{\rm rel}$, the relativistic law of velocity addition follows from the group properties.

\subsection{The rapidity space}

The RVS is also known as the ``rapidity space" \cite{Aravind1997}. Using the coordinate representation
\begin{equation}\label{eq: rapcoor}
 \fl \hskip40pt u^\mu\left( \eta,\phi,\psi \right) = \left( \cosh \eta ,\sinh\eta \sin\psi \cos\phi, \sinh\eta \sin\psi \sin\phi, \sinh\eta \cos\psi \right)
\end{equation}
in the domain $0 \le \eta < \infty, 0 \le \psi \le \pi, 0 \le \phi < 2\pi$, the line element \eref{eq: dl2} becomes
\begin{equation}\label{eq: dl2rap}
d\lambda^2 = d\eta^2 + \sinh^2\eta \left( d\psi^2 + \sin^2\psi d\phi^2 \right)
\end{equation}
For each state of motion (point in ${\cal V}_{\rm rel}$) $u^\mu = \left(\gamma(v),\gamma(v) \vec v\right)$, it follows from the relation $u^0 = \gamma \left( v \right) = \cosh\eta $ that $\eta$ is the rapidity, $\eta = \tanh^{-1}\left( v \right)$.

\subsection{Geodesics in ${\cal V}_{\rm rel}$}

The geodesic equation in ${\cal V}_{\rm rel}$ may be obtained by direct computation from the line element \eref{eq: dl2} or the metric \eref{eq: gij}. It is, however, more easily obtained by using the property that if $u^\mu\left(\lambda\right)$ is a curve on ${\cal V}_{\rm rel}$ with $\left(du/d\lambda\right)^2 = 1$ than it is a geodesic if $du^\mu/d\lambda$ is parallel transported along $u^\mu\left(\lambda\right)$ :

Any vector $A^\mu \in {\cal T}_u\left({\cal V}_{\rm rel}\right)$ satisfies $A^2 > 0 , \, A \cdot u = 0$. For any such vector, the change under parallel transport from $u^\mu$ to $u^\mu + du^\mu$ (from ${\cal T}_u\left({\cal V}_{\rm rel}\right)$ to ${\cal T}_{u+du}\left({\cal V}_{\rm rel}\right)$) is $dA^\mu = \left(A \cdot du\right) u^\mu$. The parallel transport equation is then
\begin{equation}\label{eq: partrans}
  \frac{D A^\mu}{d \lambda} = \frac{d A^\mu}{d \lambda} - \left( A \cdot \frac{d u}{d \lambda} \right) u^\mu = 0 \, .
\end{equation}
For a curve $u^\mu\left(\lambda\right)$ on ${\cal V}_{\rm rel}$ with $\left(d u / d \lambda \right)^2 = 1$, the geodesic equation is the condition of parallel transport (on $\mathcal{V}_{\rm rel}$) for $A^\mu = d u^\mu / d \lambda$,
\begin{equation} \label{eq: geodeq}
 \frac{d^2 u^\mu}{d \lambda^2} = u^\mu \, .
\end{equation}
All the geodesics on $\mathcal{V}_{\rm rel}$ may therefore be represented in the embedding space $E^{(1,3)}$ by
\begin{equation}\label{eq: geogen}
 u^\mu\left(\lambda\right) = \cosh\lambda a^\mu + \sinh\lambda b^\mu \, ,
\end{equation}
where $a^\mu$ and $b^\mu$ are constant (independent of $\lambda$) orthogonal unit 4-vectors, with
\begin{equation}\label{eq: geogenab}
 a\cdot a = -1 \, , \, b \cdot b = 1 \, , \, a \cdot b = 0 \, .
\end{equation}
If the curve $u^\mu\left(\lambda\right)$ is known, then $a^\mu$ and $b^\mu$ may be identified as
\begin{equation}\label{eq: abinit}
 a^\mu = u^\mu\left(\lambda = 0\right) \quad , \quad b^\mu = \frac{d u^\mu}{d \lambda} \left( \lambda = 0\right)
\end{equation}
$a^\mu \in \mathcal{V}_{\rm rel}$ may be regarded as the starting point of the geodesic, while $b^\mu \in {\cal T}_a\left({\cal V}_{\rm rel}\right)$ determines the initial direction. Geometrically, any two such vectors $a^\mu$ and $b^\mu$, when drawn from the origin of $E^{(1,3)}$, define a plane, and the geodesic \eref{eq: geogen} may be realized as the intersection of this plane with the hypersurface of $\mathcal{V}_{\rm rel}$ (in simile with the major circles on spheres).

\subsection{Rapidity and geodesic connection between velocity states}

Consider two velocity states $u_1^\mu $ and $u_2^\mu $. These could be the unit 4-velocities of two different particles in some instantaneous reference frame, or the unit velocity 4-vector of a particle in two instances along its world-line. In any case, they determine two points on $\mathcal{V}_{\rm rel}$, and may therefore be connected with a geodesic line. We may choose $a^\mu = u_1^\mu $ and
\begin{equation}\label{eq: }
 b^\mu = \frac{u_2^\mu + \left( u_1 \cdot u_2 \right)u_1^\mu}{\sqrt {\left( u_1 \cdot u_2 \right)^2 - 1}}
\end{equation}
so that $u_2^\mu = u^\mu\left(\lambda_o\right)$, in accordance with \eref{eq: geogen}, with
\begin{equation}\label{eq: }
 \cosh \lambda_o = - u_1 \cdot u_2 = \gamma \left( v \right) \, .
\end{equation}
$v$ is the relative speed between the two states. $\lambda_o = \tanh^{-1}v = \eta_{\rm rel}$ is therefore the relative rapidity between the two states. In other words, {\it the rapidity is the natural geodesic parameter on the velocity space}.

\subsection{Lorentz symmetry and geodesics on $\mathcal{V}_{\rm rel}$}\label{sec: Lorgeo}

As a 3-D hyperboloid of revolution embedded in a 4-D Euclidian space, the symmetry group of ${\cal V}_{\rm rel}$ is the orthochronous Lorentz group. The group actions are generated by operators of the general form
\begin{equation} \label{eq: RLrel}
 \mathcal{O} = {\omega^\mu}_\nu u^\nu \frac{\partial }{\partial u^\mu} \, .
\end{equation}
From \eref{eq: geogen} and \eref{eq: geogenab} it follows that along the geodesic
\begin{equation}\label{eq: geoabu}
 \cosh\lambda = - a\cdot u \quad , \quad \sinh\lambda = b \cdot u \, ,
\end{equation}
so the geodesics satisfy the equation
\begin{equation}\label{eq: dulambdab}
 \frac{d u^\mu}{d \lambda} = \sinh\lambda a^\mu + \cosh\lambda b^\mu = {\omega^\mu}_\nu u^\nu
\end{equation}
with
\begin{equation}\label{eq: omegab}
 {\omega^\mu}_\nu = a^\mu b_\nu - b^\mu a_\nu
\end{equation}
The geodesic \eref{eq: geogen} is therefore the integral curve generated by \eref{eq: RLrel} with ${\omega^\mu}_\nu$ given by \eref{eq: omegab}.

\noappendix

\rule{10cm}{1pt}



\begin{thebibliography}{10}


\bibitem{Hamilton1847}
Hamilton W R 1847 The hodograph, or a new method of expressing in symbolical language the Newtonian law of attraction {\it Proc. R. Irish Acad.} {\bf 3} 344-53

\bibitem{Maxwell}
Maxwell J C 1925 {\it Matter and Motion} (London: Sheldon Press) p107-9

\bibitem{Goodstein96}
Goodstein D L and Goodstein J R 1996 {\it Feynman's Lost Lecture: The Motion of Planets Around The Sun} (New York: Norton)

\bibitem{Milnor1983}
Milnor J 1983 On the geometry of the Kepler orbits {\it Am. Math. Monthly} {\bf 90} 353-65

\bibitem{Sivardiere1992}
Sivardi\'{e}re J 1992 Comments on the dynamical invariants of the Kepler and harmonic motions {\it Euro. J. Phys.} {\bf 13} 64-9

\bibitem{GonVilla.etal}
Gonz\'{a}lez-Villanueva A, Guillaum\'{\i}n-Espa\~na E, N\'{u}\~nez-Y\'{e}pez H N and Salas-Brito A L 1996 In velocity space the Kepler orbits are circular {\it Eur. J. Phys.} {\bf 17} 168-71
\nonum
Gonz\'{a}lez-Villanueva A, Guillaum\'{\i}n-Espa\~na E, Mart\'{\i}nez-Romero R P, N\'{u}\~nez-Y\'{e}pez H N and Salas-Brito A L 1998 From circular paths to elliptic orbits: a geometric approach to Kepler's motion {\it Eur. J. Phys.} {\bf 19} 431-8

\bibitem{Butikov2000}
Butikov E I 2000 The velocity hodograph for an arbitrary Keplerian motion {\it Eur. J. Phys.} {\bf 21} 1-6

\bibitem{Derbes2001}
Derbes D 2001 Reinventing the wheel: hodographic solutions of the Kepler problems {\it Am. J. Phys.} {\bf 69(4)} 481-9

\bibitem{KowenMathur2003}
Kowen M and Mathur H 2003 On Feynman's analysis of the geometry of Keplerian orbits {\it Am. J. Phys.} {\bf 71(4)} 397-401

\bibitem{Munoz2003}
Mu\~noz G 2003 Vector constants of the motion and orbits in the Coulomb/Kepler problem {\it Am. J. Phys.} {\bf 71(12)} 1292-3

\bibitem{Carinena.etal2016}
Cari\~nena J F, Ra\~nada M F and Santander M 2016 A new look at the Feynman `hodograph' approach to the Kepler first law {\it Eur. J. Phys.} {\bf 37} 025004

\bibitem{Boyer2004}
Boyer T H 2004  Unfamiliar trajectories for a relativistic particle in a Kepler or Coulomb potential {\it Am. J. Phys.} {\bf 72(8)} 992-7

\bibitem{MunozPavic2006}
Mu\~noz G and Pavic I 2006 A Hamilton-like vector for the special-relativistic Coulomb problem {\it Eur. J. Phys.} {\bf 27} 1007-18

\bibitem{Relhod}
Ben-Ya'acov U 2017 The hodograph method for relativistic Coulomb systems http://arxiv.org/abs/1701.08281

\bibitem{IARD2016}
Ben-Ya'acov U 2017 Back to epicycles -- relativistic Coulomb systems in velocity space {\it J. Phys.: conf. ser.} {\bf 845} 012010. doi: 10.1088/1742-6596/845/1/012010. Also : http://arxiv.org/abs/1701.04035.

\bibitem{RhodesSemon2004}
Rhodes J A and Semon M D 2004 Relativistic velocity space, Wigner rotation and Thomas precession {\it Am. J. Phys.} {\bf 72} (7) 943-60; doi: 10.1119/1.1652040

\bibitem{LLF5}
Landau L D and Lifshitz E M 1975 {\it The Classical Theory of Fields} (Oxford : Pergamon) Ch. 5

\bibitem{CriadoAlamo2001}
Criado C and Alamo N 2001 A link between the bounds on relativistic velocities and areas of hyperbolic triangles {\it Am. J. Phys.} {\bf 69(3)} 306-10 (2001); doi: 10.1119/1.1323963

\bibitem{Urbantke1990}
Urbantke H 1990 Physical holonomy, Thomas precession, and Clifford algebra {\it Am. J. Phys.} {\bf 58(8)} 747-50; doi: 10.1119/1.16401
\nonum
Urbantke H 1991 Erratum: "Physical holonomy, Thomas precession, and Clifford algebra" [Am. J. Phys. 58, 747-750 (1990)]
{\it Am. J. Phys.} {\bf 59} 1150; doi: 10.1119/1.16845

\bibitem{Aravind1997}
Aravind P K 1997 The Wigner angle as an anholonomy in rapidity space {\it Am. J. Phys.} {\bf 65(7)} 634-6; doi: 10.1119/1.18620

\bibitem{LLF39}
Landau L D and Lifshitz E M 1975 {\it The Classical Theory of Fields} (Oxford : Pergamon) Ch. 39

\bibitem{McIntosh71}
McIntosh H V 1971 Symmetry and degeneracy \textit{Group Theory and its Applications} Vol. 2, ed E M Loebl (New York : Academic Press) pp 75--144

\bibitem{GoldsteinLRL}
Goldstein H 1975 On the prehistory of the "Runge-Lenz" vector {\it Am. J. Phys.} {\bf 43(8)} 737-8.
\nonum
Goldstein H 1976 More on the prehistory of the Laplace or Runge-Lenz vector {\it Am. J. Phys.} {\bf 44(11)} 1123-4.

\bibitem{Goldstein.etal2000}
Goldstein  H, Poole C and Safko J 2000 {\it Classical Mechanics}  (New York: Addison-Wesley)

\bibitem{KhachidzeKhelash08}
Khachidze T T and Khelashvili A A 2008 \textit{Dynamical Symmetry of the Kepler-Coulomb Problem in Classical and Quantum Mechanics} (New-York : Nova Science)

\bibitem{Yoshida1988}
Yoshida T 1988 Rotating Laplace-Runge-Lenz vector leading to two relativistic Kepler's equations {\it Phys. Rev. A} {\bf 38(1)} 19-25

\bibitem{Stahlhofen2005a}
Stahlhofen A A 2005 Relativistic trajectories and the Runge-Lenz vector {\it Am. J. Phys.} {\bf 73(7)} 581

\bibitem{LRLgen}
Ben-Ya'acov U 2010 Laplace-Runge-Lenz symmetry in general rotationally symmetric systems {\it J. Math. Phys.} {\bf 51} 122902

\bibitem{Stephas1978}
Stephas P 1978 Relativistic two-body electrodynamnics {\it Am. J. Phys.} {\bf 46(4)} 360-5


\end{thebibliography}
 \end{document}